\documentclass[english,prl, twocolumn,superscriptaddress]{revtex4-1}
\usepackage[T1]{fontenc}
\usepackage[latin9]{inputenc}
\usepackage{color}
\usepackage{graphicx}
\usepackage{esint}

\makeatletter
 \@ifundefined{textcolor}{}
 {%
   \definecolor{BLACK}{gray}{0}
   \definecolor{WHITE}{gray}{1}
   \definecolor{RED}{rgb}{1,0,0}
   \definecolor{GREEN}{rgb}{0,1,0}
   \definecolor{BLUE}{rgb}{0,0,1}
   \definecolor{CYAN}{cmyk}{1,0,0,0}
   \definecolor{MAGENTA}{cmyk}{0,1,0,0}
   \definecolor{YELLOW}{cmyk}{0,0,1,0}
 }

\makeatother

\usepackage{babel}
\begin{document}

\title{On the Distribution of Plasmoids In High-Lundquist-Number Magnetic
Reconnection}

\author{Yi-Min Huang }

\affiliation{Center for Integrated Computation and Analysis of Reconnection and
Turbulence }

\affiliation{Center for Magnetic Self-Organization in Laboratory and Astrophysical
Plasmas}

\affiliation{Space Science Center, University of New Hampshire, Durham, NH 03824}

\author{A. Bhattacharjee}

\affiliation{Center for Integrated Computation and Analysis of Reconnection and
Turbulence }

\affiliation{Center for Magnetic Self-Organization in Laboratory and Astrophysical
Plasmas}

\affiliation{Space Science Center, University of New Hampshire, Durham, NH 03824}

\affiliation{Princeton Plasma Physics Laboratory, Princeton University, NJ 08543}
\begin{abstract}
The distribution function $f(\psi)$ of magnetic flux $\psi$ in plasmoids
formed in high-Lundquist-number current sheets is studied by means
of an analytic phenomenological model and direct numerical simulations.
The distribution function is shown to follow a power law $f(\psi)\sim\psi^{-1}$,
which differs from other recent theoretical predictions. Physical
explanations are given for the discrepant predictions of other theoretical
models. 
\end{abstract}


\maketitle
In recent years, significant advances have been made in understanding
the role of plasmoids (or secondary islands) in magnetic reconnection,
which is believed to be the underlying mechanism of energy release
for phenomena such as solar flares, magnetospheric substorms, and
sawtooth crashes in fusion plasmas\cite{ZweibelY2009}. Plasmoids
often form spontaneously in resistive magnetohydrodynamics (MHD) \cite{Biskamp1986,Lapenta2008,BhattacharjeeHYR2009,CassakSD2009},
Hall MHD \cite{ShepherdC2010,HuangBS2011}, and kinetic particle-in-cell
(PIC) \cite{DrakeSCS2006,DaughtonSK2006,DaughtonRAKYB2009} simulations
of large scale reconnection. Evidences of plasmoids have also been
found in the magnetotail and the solar atmosphere\cite{LinCF2008,LiuLWSLW2010,NishizukaTAS2010},
where they are demonstrated to play a significant role in particle
acceleration\cite{ChenBPYBIMDLKVFG2008}.

In the framework of resistive MHD, magnetic reconnection is governed
by the Lundquist number $S\equiv V_{A}L/\eta$, where $V_{A}$ is
the upstream Alfv\'en speed, $L$ is the reconnection layer length,
and $\eta$ is the resistivity. The classical Sweet-Parker theory
\cite{Sweet1958,Parker1957} assumes the existence of a stable, elongated
current sheet and yields the reconnection rate $\sim BV_{A}/\sqrt{S}$,
where $B$ is the upstream magnetic field. However, it has been shown
recently that when $S$ is above a critical value $S_{c}\sim10^{4}$,
the Sweet-Parker current sheet becomes unstable to the plasmoid instability,
with a growth rate that increases with $S$\cite{LoureiroSC2007,BhattacharjeeHYR2009}.
The reconnection layer changes to a chain of plasmoids connected by
secondary current sheets that, in turn, may become unstable again.
Eventually the reconnection layer will tend to a statistical steady
state characterized by a hierarchical structure of plasmoids \cite{ShibataT2001}.
Scaling laws of the number of plasmoids $n_{p}$, the widths $\delta$
and lengths $l$ of secondary current sheets have been deduced from
numerical simulations. These scaling laws can be understood by noting
that the process of break-up of the secondary current sheet will stop
when the local Lundquist number of a secondary current sheet drops
below $S_{c}$. Assuming that all secondary current sheets are close
to marginal stability, it can be deduced that $l\sim\eta S_{c}/V_{A}\sim LS_{c}/S$,
$\delta\sim l/\sqrt{S_{c}}\sim LS_{c}^{1/2}/S$, and $n_{p}\sim L/l\sim S/S_{c}$.
The reconnection rate may be estimated as $\eta J\sim\eta B/\delta\sim BV_{A}/\sqrt{S_{c}}$
, independent of $S$\cite{HuangB2010}.

\textcolor{black}{The discovery of the surprising scaling properties
of the plasmoid instability in the linear as well as nonlinear regimes,
and the ubiquity of the instability in collisional as well as collisionless
regimes have raised interest in seeking a statistical description
of the plasmoid dynamics in recent literature \cite{FermoDS2010,UzdenskyLS2010,FermoDSH2011,LoureiroSSU2012}.
However, existing theoretical models give conflicting predictions.
Using a heuristic argument based on self-similarity, Uzdensky }\textcolor{black}{\emph{et
al.}}\textcolor{black}{{} suggested that the distribution function $f(\psi)$
of plasmoids in terms of their magnetic fluxes $\psi$ follows a $f(\psi)\sim\psi^{-2}$
power law \cite{UzdenskyLS2010}. On the other hand, the kinetic model
of Fermo}\textcolor{black}{\emph{ et al.}}\textcolor{black}{{} \cite{FermoDS2010}
predicts a distribution function that decays exponentially in the
tail. In this Letter, we employ both kinetic models and direct numerical
simulations (DNS) of resistive MHD equations to study the distribution
of plasmoids. We first recast the heuristic argument of Uzdensky }\textcolor{black}{\emph{et
al. }}\textcolor{black}{in the form of a kinetic model, and show that
its steady-state solutions exhibit both a $f(\psi)\sim\psi^{-2}$
power-law regime and an exponential tail. This approach not only gives
a formal derivation of the $f(\psi)\sim\psi^{-2}$ power law, but
also elucidates when the the power-law regime makes a transition to
the exponential tail. However, the results of DNS show a power law
closer to $f(\psi)\sim\psi^{-1}$ than to $f(\psi)\sim\psi^{-2}$.
By careful analysis, we identify the physical causes for this deviation,
and propose a modified kinetic equation that yields solutions consistent
with the results of DNS.}

To fix ideas, we begin with a new model kinetic equation for the plasmoid
distribution function $f(\psi)$ as a function of the flux $\psi$
that yields the power-law solution obtained heuristically in \cite{UzdenskyLS2010}.
The distribution function $f(\psi)$ of the magnetic flux $\psi$
evolves in time due to the following four effects: (1) the fluxes
of plasmoids increase due to reconnection in secondary current sheets;
(2) new plasmoids are generated when secondary current sheets become
unstable; plasmoids are lost by (3) coalescence and (4) by advection
out of the reconnection layer. These effects can be encapsulated in
the equation
\begin{equation}
\frac{\partial f}{\partial t}+\alpha\frac{\partial f}{\partial\psi}=\zeta\delta(\psi)-\frac{fN}{\tau_{A}}-\frac{f}{\tau_{A}}.\label{eq:governing_eq}
\end{equation}
Here $N(\psi)\equiv\int_{\psi}^{\infty}f(\psi')d\psi'$ is the cumulative
distribution function, i.e. the number of plasmoids with fluxes larger
than $\psi$. In Eq. (\ref{eq:governing_eq}), the following assumptions
have been made: (1) All secondary current sheets are close to marginal
stability, therefore on average all plasmoids grow at a constant rate
$\alpha\sim BV_{A}/\sqrt{S_{c}}$. (2) When new plasmoids are created,
they contain zero flux (represented by the source term $\zeta\delta(\psi$),
where $\delta(\psi)$ is the Dirac $\delta-$function). (3) Plasmoids
disappear upon encountering larger plasmoids. This is represented
by the loss term $-fN/\tau_{A}$, where the characteristic time scale
to encounter a larger plasmoid is estimated as $\sim\tau_{A}/N\equiv L/NV_{A}$,
assuming the characteristic relative velocity between plasmoids is
of the order of $V_{A}$. The process of coalescence is assumed to
be instantaneous. Note that when two plasmoids coalesce, the flux
of the merged plasmoid is equal to the larger of the two original
fluxes \cite{FermoDS2010}. Therefore, coalescence does not affect
the value of $f$ at the larger of the two fluxes. (4) Lastly, plasmoid
loss due to advection is represented by the term $-f/\tau_{A}$, where
the time scale $\tau_{A}$ is based on the outflow speed $\sim V_{A}$. 

\begin{figure}[t]
\begin{centering}
\includegraphics[scale=0.8]{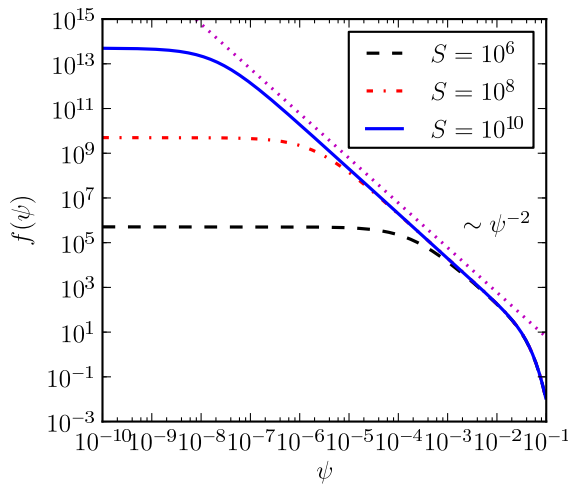}
\par\end{centering}

\caption{(Color online) The distribution function (\ref{eq:distribution})
for $S=10^{6}$, $10^{8}$, and $10^{10}$. \label{fig:Distribution-psim2}}
\end{figure}

Under steady-state conditions, Eq. (\ref{eq:governing_eq}) admits
the analytic solution

\begin{equation}
f(\psi)=\frac{2C/\alpha\tau_{A}}{\left(C-\exp(-\psi/\alpha\tau_{A})\right)^{2}}\exp(-\psi/\alpha\tau_{A}),\label{eq:distribution}
\end{equation}
where the constant $C=1+2/n_{p}$ , with the total number of plasmoids
$n_{p}=\int_{0}^{\infty}f(\psi)d\psi$. The source term $\zeta\delta(\psi)$
sets the boundary condition $f(0)=\zeta/\alpha$, which gives the
relation $\zeta\tau_{A}=n_{p}^{2}/2+n_{p}.$ The source term magnitude
$\zeta$ may be estimated by the relation $n_{p}\sim S/S_{c}$. In
the limit $S\gg S_{c}$, we have $\zeta\sim n_{p}^{2}/2\tau_{A}\sim(S/S_{c})^{2}/2\tau_{A}$.
The distribution function (\ref{eq:distribution}) has three distinct
regimes when $S\gg S_{c}$ : (i) $f\simeq(2/\alpha\tau_{A})\exp(-\psi/\alpha\tau_{A})$
when $\psi/\alpha\tau_{A}\gg1$; (ii) $f\simeq2\alpha\tau_{A}\psi^{-2}$
when $2/n_{p}\ll\psi/\alpha\tau_{A}\ll1$; (iii) $f\simeq n_{p}^{2}/2\alpha\tau_{A}$
when $\psi/\alpha\tau_{A}\ll2/n_{p}$. Therefore, the solution admits
both an exponential tail and a power-law regime. It can be shown that
the dominant loss mechanism in the former regime is advection ($N\ll1$;$\alpha\partial f/\partial\psi\simeq-f/\tau_{A}$),
while it is coalescence in the latter ($N\gg1$; $\alpha\partial f/\partial\psi\simeq-fN/\tau_{A}$).
Figure \ref{fig:Distribution-psim2} shows the distribution function
(\ref{eq:distribution}) for $S=10^{6}$, $10^{8}$, and $10^{10}$.
Here, to fix ideas, we have taken $S_{c}=10^{4}$, $V_{A}=1$, $B=1$,
and $L=1$, and all scaling relations, such as $\alpha\sim BV_{A}/\sqrt{S_{c}}$,
are replaced by equalities. Note that the range where the $f\sim\psi^{-2}$
power law holds is more extended for higher $S$.

\begin{figure}[t]
\begin{centering}
\includegraphics[scale=0.8]{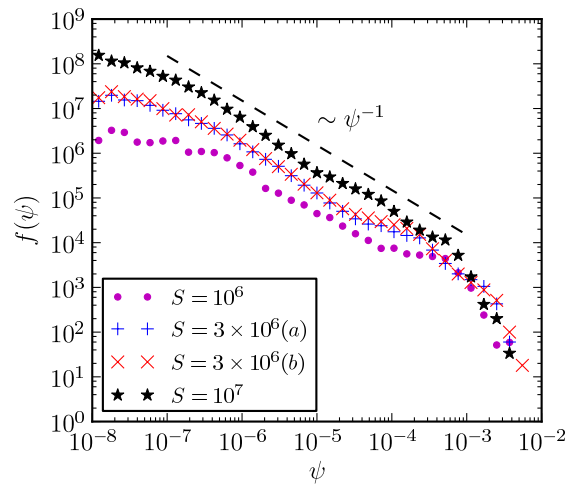}
\par\end{centering}

\caption{(Color online) Plasmoid distribution functions from direct numerical
simulations. \label{fig:Plasmoid-distributions-from-DNS}}
\end{figure}

To test the $f(\psi)\sim\psi^{-2}$ power law by DNS, we use the same
simulation setup of two coalescing magnetic islands as in a previous
study \cite{HuangB2010}.\textcolor{black}{{} }The 2D simulation box
is \textcolor{black}{the domain $(x,z)\in[-1/2,1/2]\times[-1/2,1/2]$}.
In normalized units, the initial magnetic field \textcolor{black}{is
given by $\mathbf{B}_{0}=\nabla\psi_{0}\times\mathbf{\hat{y}}$, where
}$\psi_{0}=\tanh\left(z/h\right)\cos\left(\pi x\right)\sin\left(2\pi z\right)/2\pi$.
The parameter $h$, which is set to $0.01$ for all simulations, determines
the initial current layer width. The initial plasma density $\rho$
is approximately $1$, and the plasma temperature $T$ is $3$. The
density profile has a weak nonuniformity such that the initial condition
is approximately force-balanced. The initial peak magnetic field and
Alfv\'en speed are both approximately unity. The plasma beta $\beta\equiv p/B^{2}=2\rho T/B^{2}$
is greater than $6$ everywhere. Perfectly conducting and free slipping
boundary conditions are imposed along both $x$ and $z$ directions.
\textcolor{black}{Only the upper half of the domain ($z\ge0$) is
simulated, and solutions in the lower half are inferred by symmetries.
We use a uniform grid along the $x$ direction and a nonuniform grid
along the $z$ direction that packs high resolution around $z=0$.
For cases with $S=10^{6}$ and $3\times10^{6}$, the mesh size is
$12726\times1600$, and the smallest grid size along $z$ is $5.7\times10^{-6}$.
For the $S=10^{7}$ case, the mesh size is $37800\times2880$, and
the smallest grid size along $z$ is $1.9\times10^{-6}$. No explicit
viscosity is employed in these simulations. A fourth order numerical
dissipation is added to damp small fluctuations at grid scale \cite{GuzdarDMHL1993}. }

\textcolor{black}{The initial velocity is seeded with a random noise
of amplitude $10^{-6}$ to trigger the plasmoid instability. The early
period when the reconnected flux is less than 0.01 is precluded from
the analysis to allow the reconnection layer to reach a statistical
steady state. We take data during the period when the reconnected
flux is between $0.01$ to $0.05$, corresponding to $25\%$ of the
initial flux in each of the merging islands. This period roughly spans
$6\tau_{A}$, insensitive to $S$. Snapshots are taken at intervals
of $0.01\tau_{A}$. We identify plasmoids within the range $x\in[-0.25,0.25]$
with a computer program for each snapshot, which provide the dataset
for further statistical analysis. }Figure \ref{fig:Plasmoid-distributions-from-DNS}
shows the probability distribution functions $f(\psi)$ for $S=10^{6}$,
$3\times10^{6}$ {[}two runs, labeled as (a) and (b){]}, and $S=10^{7}$.
Distribution functions are normalized such that $\int_{0}^{\infty}f(\psi)d\psi$
is equal to the average number of plasmoids in each time slice. These
numerical results appear to be robust and reproducible, as exemplified
by the two $S=3\times10^{6}$ runs that yield nearly identical distribution
functions. Qualitative similarities between Fig. \ref{fig:Distribution-psim2}
 and Fig. \ref{fig:Plasmoid-distributions-from-DNS}, especially the
existence of three distinct regimes, are evident. However, the distribution
function in the power-law regime is closer to $f(\psi)\sim\psi^{-1}$
instead of $f(\psi)\sim\psi^{-2}$. 

\begin{figure}
\begin{centering}
\includegraphics[scale=0.8]{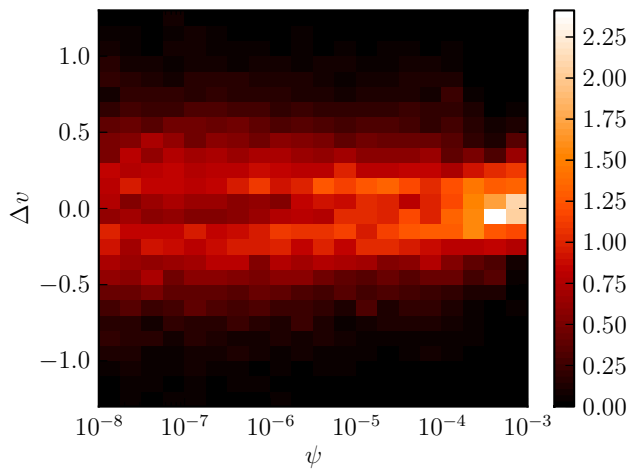}
\par\end{centering}

\caption{(Color online) The plasmoid distribution with respect to the relative
speed $\mbox{\ensuremath{\Delta}}v$ and the flux $\psi$ from the
run $S=10^{7}$. \label{fig:Relative-velocity}}

\end{figure}

To understand the discrepancy between the numerical results and the
power-law prediction of Eq. (2), we need to critically examine the
basic assumptions that give rise to the $f(\psi)\sim\psi^{-2}$ power
law. In the $f(\psi)\sim\psi^{-2}$ regime, the dominant balance in
Eq. (\ref{eq:governing_eq}) is between the plasmoid growth term and
the loss term due to coalescence, i.e. $\alpha\partial f/\partial\psi\simeq-fN/\tau_{A}$.
A key assumption underlying the loss term $-fN/\tau_{A}$ is that
the relative speeds of a plasmoid with respect to neighboring plasmoids
larger than itself are of the order of $V_{A}$ and are uncorrelated
to the flux of the plasmoid. To examine this assumption with numerical
data, we measure the relative velocity $\Delta v$ of each plasmoid
at any given time with respect to the first larger plasmoid it will
encounter by extrapolating the trajectories of the plasmoids with
their velocities at that time. Note that $\Delta v$ is undefined
for the largest plasmoid, or when all larger plasmoids are moving
away from a given plasmoid. The plasmoids with $\Delta v$ undefined
are disregarded in the analyses. Figure \ref{fig:Relative-velocity}
shows the distribution $g(\psi,\Delta v)$ of plasmoids with respect
to $\psi$ and $\Delta v$ from the run $S=10^{7}$. Here we normalize
$g(\psi,\Delta v)$ such that $\int_{-\infty}^{\infty}g(\psi,\Delta v)d(\Delta v)=1$
for better visualization. We can clearly see that the distribution
is not uniform across different values of $\psi$. The distribution
covers a broader range of $\Delta v$ at smaller $\psi$, and it becomes
more concentrated around $\Delta v=0$ at larger $\psi$. Similar
results are also observed in other runs. Therefore, it appears that
the reconnection layer organizes itself spontaneously into a state
such that large plasmoids tend to avoid coalescing with each other. 

\begin{figure}
\begin{centering}
\includegraphics[scale=0.8]{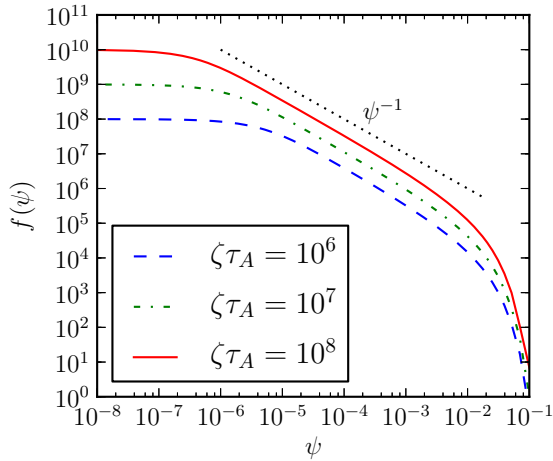}
\par\end{centering}

\caption{(Color online) Distribution functions from numerical solutions of
Eq. (\ref{eq:modified}). \label{fig:Modified_sol}}
\end{figure}

How do we interpret this phenomenon? As discussed earlier, the flux
of a plasmoid is approximately proportional to its age because all
plasmoids grow approximately at the same rate $\alpha$. Consequently,
a plasmoid can become large only if it has not encountered plasmoids
larger than itself for an extended period of time. Presumably, plasmoids
moving rapidly relative to their neighbors will encounter larger plasmoids
and disappear easily, whereas those with small relative speeds are
more likely to survive for a long time and become large. This observation
motivates us to consider a distribution function $F(\psi,v)$, where
$v$ can be interpreted as the plasmoid velocity relative to the mean
flow (which has a profile along the outflow direction). The governing
equation for $F(\psi,v)$ is written as 
\begin{equation}
\partial_{t}F+\alpha\frac{\partial F}{\partial\psi}=\zeta\delta(\psi)h(v)-\frac{FH}{\tau_{A}}-\frac{F}{\tau_{A}},\label{eq:modified}
\end{equation}
where the function $H$ is defined as 
\begin{equation}
H(\psi,v)=\int_{\psi}^{\infty}d\psi^{'}\int_{-\infty}^{\infty}dv'\frac{\left|v-v'\right|}{V_{A}}F(\psi',v'),\label{eq:collision}
\end{equation}
and $h(v)$ is an arbitrary distribution function in velocity space
when new plasmoids are generated. The distribution function $f(\psi)$
can be obtained by integrating $F(\psi,v)$ over the velocity space.
Eq. (\ref{eq:modified}) differs from Eq. (\ref{eq:governing_eq})
in the plasmoid loss term due to coalescence, where the relative speed
$\left|v-v'\right|$ between two plasmoids is taken into account in
the integral operator of Eq. (\ref{eq:collision}). If we replace
$\left|v-v'\right|$ in Eq. (\ref{eq:collision}) by $V_{A}$, then
Eq. (\ref{eq:modified}) reduces to Eq. (\ref{eq:governing_eq}).
Steady-state solutions of Eq. (\ref{eq:modified}) can be obtained
numerically. To fix ideas, we assume a Gaussian profile $h(v)=(1/\sqrt{\pi}V_{A})\exp(-v^{2}/V_{A}^{2})$
for the arbitrary source function. Fig. \ref{fig:Modified_sol} shows
the resulting $f(\psi)$ for $\zeta\tau_{A}=10^{6},$ $10^{7}$, and
$10^{8}$. Assuming $n_{p}\simeq S/S_{c}$ and $S_{c}\simeq10^{4}$,
these solutions approximately correspond to $S=3\times10^{7}$, $10^{8}$,
and $3\times10^{8}$, respectively. These solutions also show three
distinct regimes as the solutions in Fig. \ref{fig:Distribution-psim2}.
However, the distribution in the intermediate power-law regime is
close to $f(\psi)\sim\psi^{-1}$, consistent with DNS. We have tried
other smooth $h(v)$ profiles, and the results do not appear to be
sensitive to the specific form of $h(v)$, as long as $h(v)$ covers
a broad range of $v$ (typically of the order of $V_{A}$).

\textcolor{black}{A previous DNS study of plasmoid distribution has
been recently carried out by Loureiro }\textcolor{black}{\emph{et
al.}}\textcolor{black}{{} \cite{LoureiroSSU2012}, where they claimed
confirmation of the $f(\psi)\sim\psi^{-2}$ distribution. It should
be pointed out that Loureiro}\textcolor{black}{\emph{ et al. }}\textcolor{black}{compared
the $f(\psi)\sim\psi^{-2}$ prediction with simulation data in the
large-$\psi$ regime. If we focus on the smaller-$\psi$ regime of
their numerical data, the distribution appears more consistent with
our finding $f(\psi)\sim\psi^{-1}$. This flattening of distribution
function in the smaller-$\psi$ regime was noted by Loureiro}\textcolor{black}{\emph{
et al., }}\textcolor{black}{but no attempt was made to fit the smaller-$\psi$
regime to a power law. An important question is: do we expect to see
a power law in the large-$\psi$ regime or the smaller-$\psi$ regime?
Our analytic theory reveals that the transition from a power-law distribution
to an exponential tail is due to a change in the dominant loss mechanism
from coalescence to advection, which occurs approximately when $N\sim O(1)$.
In our simulation data, the cumulative distribution function $N(\psi)$
drops below unity at $\psi\sim10^{-3}$, which is also approximately
where the distribution function deviates from $f(\psi)\sim\psi^{-1}$
to a more rapid, presumably exponential, falloff. Therefore, this
rapidly falling tail is not where a power law should arise. However,
simulation data in the large-$\psi$ regime is sufficiently uncertain
that it may be difficult to make a clear distinction between a $\psi^{-2}$
and an exponential falloff. Note that the exponential falloff at large
$\psi$ is consistent with the prediction of the kinetic model of
Fermo}\textcolor{black}{\emph{ et al.}}\textcolor{black}{{} \cite{FermoDS2010}
and a subsequent analysis of the flux transfer events (FTEs) in the
magnetopause from Cluster \cite{FermoDSH2011}. Fermo}\textcolor{black}{\emph{
et al.}}\textcolor{black}{{} did not explicitly address the distribution
of smaller plasmoids. Because the coalescence term in their model
is based on very different considerations and assumptions, it is not
clear whether the distribution of smaller plasmoids will follow a
power law.}

\textcolor{black}{Although Eq. (\ref{eq:modified}) is a significant
improvement on Eq. (\ref{eq:governing_eq}), it does not include some
important physical effects. Most notably, coalescence between islands
is assumed to occur instantaneously, whereas in reality larger plasmoids
take longer to merge, and there can be bouncing (or sloshing) between
them \cite{KnollC2006,KarimabadiDRDC2011}. These effects may also
contribute to the distribution shown in Fig. \ref{fig:Relative-velocity}.
Furthermore, the velocity $v$ relative to the mean flow is assumed
to remain constant throughout the lifetime of a plasmoid, whereas
in reality some variation is expected due to the complex dynamics
between plasmoids. Finally, in high-$S$ regime the current sheet
between two coalescing plasmoids can also be the source of more plasmoids.\cite{BartaBKS2011}.}

\textcolor{black}{It should be borne in mind that our considerations
are valid for collisional plasmas obeying the resistive MHD equations.
In weakly collisional systems the plasmoid instability inevitably
drives reconnection towards the collisionless regime \cite{DaughtonRAKYB2009,ShepherdC2010,HuangBS2011}.
The question of the plasmoid distribution in the collisionless regime
remains largely open. However, some of the key ideas in this work,
such as the tendency of large plasmoids to avoid coalescence, may
still be relevant. }The present study is limited to highly idealized
2D problems where more concrete conclusions can be drawn. In 3D geometry
oblique tearing modes have been shown to play an important role \cite{DaughtonRKYABB2011,BaalrudBH2012},
and a statistical description of such systems remains a great challenge. 

This work was supported \textcolor{black}{by the Department of Energy,
Grant No. DE-FG02-07ER46372, under the auspice of the Center for Integrated
Computation and Analysis of Reconnection and Turbulence (CICART),
the National Science Foundation, Grant No. PHY-0215581 (PFC: Center
for Magnetic Self-Organization in Laboratory and Astrophysical Plasmas),
NASA Grant Nos. NNX09AJ86G and NNX10AC04G, and NSF Grant Nos. ATM-0802727,
ATM-090315 and AGS-0962698. Computations were performed on Oak Ridge
Leadership Computing Facility through an INCITE award, and National
Energy Research Scientific Computing Center.}

\bibliographystyle{apsrev4-1}

\end{document}